\documentclass{article}
\pdfoutput=1 
\newcommand{\TheTitle}{An Exact Stochastic Simulation Method for Fractional Order Compartment Models}

\usepackage{amsmath}
\DeclareMathOperator*{\argmin}{arg\,min}
\usepackage{amsfonts}
\usepackage{graphicx}
\usepackage{caption}
\usepackage{subcaption}
\graphicspath{{Figures/}} 

\providecommand{\msc}[1]{\textbf{MSC:} #1}

\title{{\TheTitle}}

\author{
  Christopher N. Angstmann\thanks{School of Mathematics and Statistics, UNSW Australia Sydney, NSW 2052, Australia} \and Stuart-James Burney\footnotemark[1] \and Bruce I. Henry\footnotemark[1] \and Daniel S. Han\footnotemark[1] \and Byron A. Jacobs\thanks{
    	School of Computer Science and Applied Mathematics, 
    	University of the Witwatersrand, 
    	Johannesburg, 
    	Private Bag 3, 
    	Wits 2050, 
    	South Africa} \and Zhuang Xu\footnotemark[1] 
     }

\usepackage{amsopn}

\begin{document}

	\maketitle
	
	\begin{abstract}
	    Our study focuses on fractional order compartment models derived from underlying physical stochastic processes, providing a more physically grounded approach compared to models that use the dynamical system approach by simply replacing integer-order derivatives with fractional order derivatives. In these models, inherent stochasticity becomes important, particularly when dealing with the dynamics of small populations far from the continuum limit of large particle numbers. The necessity for stochastic simulations arises from deviations of the mean states from those obtained from the governing equations in these scenarios. To address this, we introduce an exact stochastic simulation algorithm designed for fractional order compartment models, based on a semi-Markov process. We have considered a fractional order resusceptibility SIS model and a fractional order recovery SIR model as illustrative examples, highlighting significant disparities between deterministic and stochastic dynamics when the total population is small. Beyond its modeling applications, the algorithm presented serves as a versatile tool for solving fractional order differential equations via Monte Carlo simulations.
		
	\end{abstract}
	

	\msc{34A08, 60G22, 60K40, 92C45, 92D30}
	
	\section{Introduction}
	Compartmental models have gained widespread use across various mathematical modelling fields, including epidemiology, pharmacokinetics, and chemical kinetics \cite{Savic2007,Tsoukias1998,Marino2011,Olson2001,Kermack1927}. Traditionally, these models have relied on the assumption that the time to the next event or inter-compartmental interaction follows an exponential distribution. However, more recent research has shifted its focus toward models that incorporate non-Markovian dynamics \cite{VanNonMarkovianSIS2013, IstvanNonMarkovianEpi2015, SaeedianMemoryEpi2017, ShkilevNonMarkovCompartment2019}. In these models, the time to the next inter-compartmental interaction follows a heavy-tailed distribution, which allows for the possibility of significantly prolonged waiting times between events. Intriguingly, non-Markovian reactions, when coupled with diffusion, have been shown to induce deviations in dynamics that are in contradiction with the classical picture of law of mass action \cite{Campos2008}.
	
	Typically, a system of differential equations is employed to analyze the mean behavior of modelled systems. When addressing the influence of environmental changes, stochasticity can be incorporated into epidemiological models through the introduction of a random noise process. However, such introductions are often performed in an {\em ad-hoc} manner and may result in physically unrealistic negative solutions \cite{GUO2017,DU2017,Anqi2018}. Alternatively, more grounded approaches introduce noise into the  transmission rate rather than directly into the population \cite{JiRatePerturbSIR2011,KangSISLevy2017}. 
	On the other hand, intrinsic stochasticity naturally emerges as an important factor in the context of small population dynamics.
	
	The Gillespie algorithm \cite{GillespieDanielT1976Agmf,GillespieDanielT1977Exact} offers a general framework for stochastically simulating chemical reactions and compartmental models, enabling the generation of a single path-wise realization of the ensemble process. 
	In cases involving sufficiently large populations, stochastic and deterministic models both provide adequate descriptions of the mean dynamical behavior.
	However, when dealing with scenarios deviating from this limiting case, conventional deterministic differential equations struggle to accurately represent small population dynamics. This is especially important when studying extinction events \cite{ALLEN2000,ALLEN2012}, and commonly referred to the ``atto-fox" problem in the context of predator-prey models \cite{CAMPILLO2012}.
	
	The extension of the Gillespie algorithm to non-Markovian processes has been explored previously \cite{BogunaNonMarkovStochastic2014, VesNonMarkovProcess2015, NaokiLaplaceGillespie2018}. Adapting these algorithms to accommodate general fractional order compartment models, where particles may experience both Markovian and non-Markovian dynamics in a single compartment, poses a unique challenge. To address this, we introduce an exact stochastic simulation algorithm that integrates elements from the Gillespie algorithm \cite{GillespieDanielT1977Exact} and the Next Reaction Method \cite{MichaelNextReactionMethod2000}. This hybrid approach offers versatile modeling capabilities for fractional order compartment models. When waiting times between transitions of particles for the non-Markovian dynamics are drawn from the Mittag-Leffler waiting time distributions, the corresponding governing differential equations will incorporate Riemann-Louiville fractional derivatives. In this work, we will highlight the relationship between stochastic and deterministic descriptions of the fractional order compartment models in the continuum limit of large particle numbers, and their divergent behaviors when departing from this limit. To elucidate these concepts and showcase the effectiveness of our algorithm, we investigate epidemiological models, specifically a fractional order resusceptibility susceptible-infected-susceptible (SIS) model and a fractional order recovery susceptible-infected-recovered (SIR) model, as practical examples.
	\\
	
	\section{ The Stochastic Process}
	
\begin{figure}[htbp]
		\begin{center}
			\includegraphics[width=0.7\linewidth]{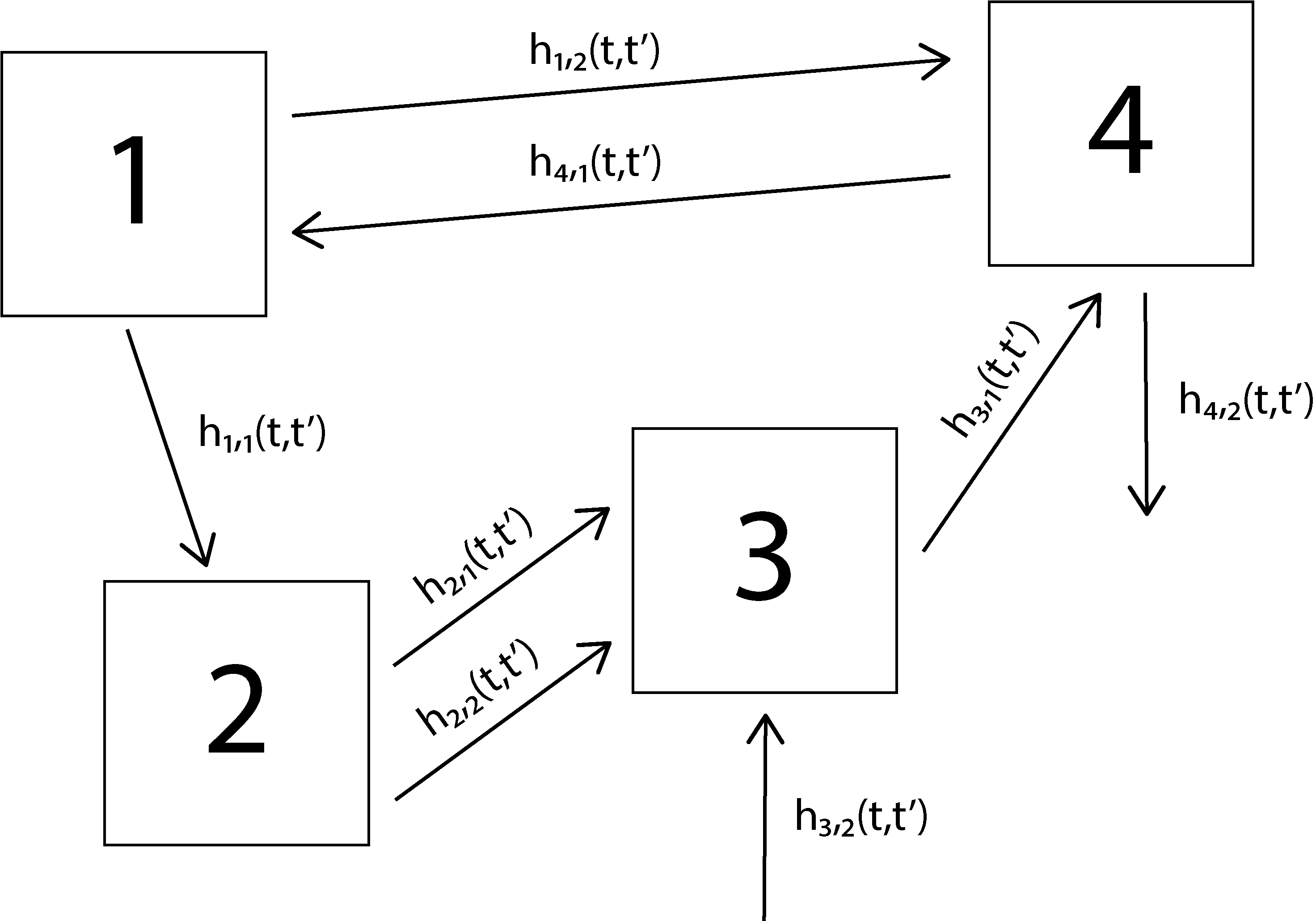}
			\caption{A box and arrow diagram for a non-Markovian compartment model. The arrows are annotated with the per-particle transition rate, denoted as $h_{i,j}(t,t')$, which may depend on both the current time, $t$, and the time, $t'$, when the particle arrived in the compartment. Here, $i$ and $j$ represent the originating compartment for the removal process and the process label, respectively. In cases where the transition rate does not originate from a specific compartment (e.g., $h_{3,2}(t,t')$), such as a population-independent birth process, $i$ denotes the compartment to which the process arrives.}
			\label{fig_BA}
			\vspace{-1.5\baselineskip}
		\end{center}
	\end{figure}

	Fractional order compartment models may be derived from an underlying stochastic process \cite{AngstmannFOCM2017}. 
	We begin by considering a non-Markovian compartment model, where particles reside in each compartment for a finite time before transitioning between compartments instantaneously. 
	We make the assumption that, for each particle, the transition rate between compartments may depend on both the current time, $t$, and the time the particle entered the compartment $t'$. In order to arrive at a fractional order compartment model, we additionally assume that each compartment is subject to a single non-Markovian removal process alongside an arbitrary number of Markovian processes. The set of such rates and compartments will completely define the stochastic process. Due to the non-Markovian nature of the model, two particles in the same compartment are subject to different rates depending on the times they arrived. An illustrative box and arrow diagram is given in Figure \ref{fig_BA}. 
	
	In a model comprising of $N$ compartments, the system's state at time $t$ is fully characterized by the compartment in which the particles reside and the time at which they entered their current compartment. This can be represented using $N$ vectors, $\mathbf{t'}$, each with length $M_i(t)$, where the elements of the vectors correspond to the arrival time of a particle. The length $M_i(t)$ signifies the number of particles in compartment $i$ at time $t$. In this way, the total number of particles in the system at time $t$ is given by
	\begin{equation}
		T(t)=\sum_{i=1}^{N}M_{i}(t).
	\end{equation}
	Consequently, our description of the system state is
	\begin{equation}
		X(t)=\{\mathbf{t'}_{M_1(t)},\mathbf{t'}_{M_2(t)},\mathbf{t'}_{M_3(t)},\dots,\mathbf{t'}_{M_N(t)}\}.
	\end{equation} 
	Here $\mathbf{t'}_{M_i(t)}$ denotes the vector of arrival times $t'$ of length $M_i(t)$ for compartment $i$. The system's state will change every time a particle transitions between compartments. 
	
	\section{Mean Governing Equations}
	
	Here we construct the mean governing equations for this stochastic process.  In general this mean is formed in a continuum limit of particle numbers, where we make the assumption that the number of particles is sufficiently large so that the fluctuations of the process are negligible.  
	
	\subsection{Integro-Differential Equation Description} The evolution of the mean state of the stochastic compartment dynamics can be described via a set of integro-differential equations. First, we begin by considering the probability that a transition event associated with the removal process $j$ in compartment $i$, characterized by a rate of $h_{i,j}(t,t')$, will not happen before time $t$, given that it entered the compartment at time $t'$. This probability can be expressed as
	\begin{equation}
		\Phi_{i,j}(t,t')=\exp\left({-\int_{t'}^{t}h_{i,j}(z,t')dz}\right).
	\end{equation}
	This is commonly referred to as the survival function. We assume that the rates of individual processes within the compartment are independent. In this case, we can express the total survival function for compartment $i$ as a product of the survival functions for each process, given as 
	\begin{equation}
		\Phi_i(t,t')=\prod_{j}\Phi_{i,j}(t,t').
	\end{equation} 
	The expected number of particles in compartment $i$ at time $t$, 
	can be expressed in terms of $q_i(t)$,  the expected arrival flux of particles into compartment $i$ at time $t$, and the survival probability $\Phi_i$ as 
	\begin{equation}\label{eq_U2}
		u_i(t)=\int_{0}^{t}\Phi_i(t,t')q_i(t')dt'.
	\end{equation} 
	
	To derive an integro-differential governing equation, we will consider a restricted case where compartment $i$ has $n_i$ rates ($\mu_{i,j}(t)$, $j\in\{1,2,...,n_i\}$) that solely depend on the current time (Markovian), and a single rate $\gamma_i$ that depends on the time since arrival (non-Markovian). With these assumptions, we can express the total survival function as
	\begin{equation}
		\begin{split}
			\Phi_i(t,t') &=\exp\left({-\int_{t'}^{t}\gamma_{i}(z-t')dz}\right)\exp\left({-\int_{t'}^{t}\sum_{j=1}^{n_i}\mu_{i,j}(z)dz}\right)\\
			&= \Psi_i(t-t')\Theta_i(t,t'),
		\end{split}
	\end{equation} 
	where $\Psi_i(t-t')$ and $\Theta_i(t,t')$ denote the survival probabilities for the non-Markovian removal process and all the associated Markovian processes in compartment $i$, respectively. Next, we differentiate \eqref{eq_U2} with respect to time $t$, provided that $q_i(t)$ is continuous for $t\geq0$, which yields
	\begin{equation}\label{eq_dU1}
		\frac{du_i(t)}{dt}=q_i(t)-\int_{0}^{t}\phi_i(t,t')q_i(t')dt',
	\end{equation} 
	where 
	\begin{equation}
		\phi_i(t,t')=-\frac{d \Phi_i(t,t')}{dt}.
	\end{equation}
	In terms of the rates, this gives
	\begin{equation}\label{eq_dU3}
		\begin{split}
			\frac{du_i(t)}{dt}=&q_i(t)-\sum_{j=1}^{n_i}\mu_{i,j}(t)u_{i}(t)-\int_{0}^{t}\gamma_{i}(t-t')\Phi_i(t,t')q_i(t')dt'.
		\end{split}
	\end{equation}
	Let $\psi_i(t)$  denote the waiting time distribution for the non-Markovian removal process, such that 
	\begin{equation}
		\frac{d\Psi_{i}(t)}{dt}=-\psi_{i}(t).
	\end{equation}
	Using the Laplace transform approach as demonstrated in \cite{AngstmannFOCM2017}, the governing equation for compartment $i$ is given by
	\begin{equation} \label{eq_dU4}
		\begin{split}
			\frac{du_i(t)}{dt}=&q_i(t)-\omega_i(t)u_{i}(t)-\int_{0}^{t}K_i(t-t')\Theta_i(t,t')u_i(t')dt',
		\end{split}
	\end{equation}
	where $\omega_i(t)=\sum_{j=1}^{n_i}\mu_{i,j}(t)$ represents the total rate for all the associated Markovian processes in the compartment and the memory kernel, $K_i(t)$, is defined via the Laplace-transform as
	\begin{equation} \label{eq_dU5}
		\begin{split}
			\mathcal{L}_t\{K_i(t)\} = \frac{\mathcal{L}_t\{\psi_i(t)\}}{\mathcal{L}_t\{\Psi_i(t)\}}.
		\end{split}
	\end{equation}
	Note that in the situation where there is an initial injection of flux arriving at the compartment $i$ at $t=0$, we can write the arrival flux $q_i(t)$ as
	\begin{equation} \label{eq_i0flux}
		q_i(t) = i_0\delta(t-0^+)+q_i^+(t),
	\end{equation}
	where $i_0$ is the initial injection in the compartment and $q_i^+(t)$ is right continuous at $t=0$ and continuous for $t>0$. In this case, the governing equation, \eqref{eq_dU4}, will be
	\begin{equation} \label{eq_dU6}
		\frac{du_i(t)}{dt}=q^+_i(t)-\omega_i(t)u_{i}(t)-\int_{0}^{t}K_i(t-t')\Theta_i(t,t')u_i(t')dt'.
	\end{equation}
	For a model consisting of $N$ compartments, the dynamics of the system's mean state are described by a set of integro-differential equations (\ref{eq_dU4}), for $i=1,2,...,N$. In general, the arrival rate into each compartment may depend on the expected number of particles leaving other compartments. This is achieved by matching the arrival fluxes to the removal fluxes from the other compartments.  \\

	\subsection{ Fractional Order Equation Description}
	The fractional derivative arises when the non-Markovian removal process has a power-law tailed waiting time distribution. In this scenario, the rate of removal decreases as particles have resided in the compartment for a longer duration. The Mittag-Leffler waiting time distribution is one of the power-law tailed distributions with an asymptotic decay $\psi_{i}(t)\sim t^{-\alpha-1}$ as $t\rightarrow{\infty}$. The corresponding survival function can be expressed in terms of a Mittag--Leffler function as
	\begin{equation}\label{eq_MLsurvival}
		\Psi_{i}(t) = E_{\alpha_i,1}\left(-\left(\frac{t}{\tau_i}\right)^{\alpha_i}\right)
	\end{equation}
	with the exponent $0<\alpha_i\leq1$ and the time scale parameter $\tau_i>0$. Here $E_{\alpha,\beta}(t)$ is the two parameter Mittag-Leffler function, defined by
	\begin{equation}\label{eq_MLDefinition}
		E_{\alpha,\beta}(t)=\sum_{k=0}^\infty\frac{t^k}{\Gamma(\alpha k+\beta)}, \quad \beta\in\mathbb{C}.
	\end{equation}
	If we take the non-Markovian waiting time to be Mittag-Leffler distributed in \eqref{eq_dU4}, then the governing evolution equation can be written as
	\begin{equation}\label{eq_GM2}
		\begin{split}
			\frac{du_i(t)}{dt}=&q_i(t)-\omega_i(t)u_{i}(t)-\tau_i^{-\alpha_i}\Theta_i(t,0)_{0}\mathcal{D}_t^{1-\alpha_i}\left\{\frac{u_i(t)}{\Theta_i(t,0)}\right\},
		\end{split}
	\end{equation}
	where $_{0}\mathcal{D}_t^{1-\alpha}\{f(t)\}$ is the Riemann--Liouville fractional derivative of order $1 - \alpha$, defined by
	\begin{equation}\label{eq_RLDericative}
		_{0}\mathcal{D}_t^{1-\alpha}f(t)=\frac{1}{\Gamma(\alpha)}\frac{d}{dt}\int_{0}^{t}\frac{f(t')}{(t-t')^{1-\alpha}}dt'.
	\end{equation}
	Other power-law tailed waiting time distributions, such as the Pareto distribution, can also be considered and will lead to a similar equation as \eqref{eq_GM2} in the asymptotic limit as $t\rightarrow\infty$. However, the choice of the Mittag-Leffler waiting time ensures the validity of the governing equation for all $t\ge0$.
	
	\section{Stochastic Simulation Method}
	The simulation of the stochastic process corresponding to the fractional order compartment model can be achieved by drawing the appropriately distributed waiting times. The state of the stochastic system is completely described by the number of particles in each compartment and the time at which each of the particles arrived in their current compartment. If a compartment only contains Markovian processes, then the individual arrival times can be effectively represented by the number of particles within the compartment using the Gillespie algorithm.
	
	The simulation algorithm presented here is a hybrid approach, combining elements of the Gillespie algorithm \cite{GillespieDanielT1977Exact} and the Next Reaction Method \cite{MichaelNextReactionMethod2000}. This algorithm drives the evolution of a system's state by generating a time to the next transition for each compartment. Instead of tracking the arrival time of each particle in a compartment, we have chosen to store the time remaining until the particle leaves the compartment. This alternative representation provides an equivalent description of the stochastic system.
	
	We consider a general compartment model with $N$ compartments. We will categorise each compartment as either Markovian or non-Markovian, depending on if the rates leaving the compartment are all Markovian, or if there exists a non-Markovian removal process. Note that the non-Markovian compartment may also contain Markovian processes.
	
	Firstly we will consider the Markovian compartments. As each of the particles in the compartment are indistinguishable, the time to next event within this compartment is completely defined by the number of particles currently in the compartment and the rates of all the Markovian processes. Thus the evolution algorithm will store three numbers for each Markovian compartment: the time to the next particle transition, the label of the next transition process and the current number of particles in the compartment. The time to the next transition can be generated via the transition rates. For a Markovian compartment $i$ with $n_i$ possible transitions and $M_i(t)$ particles at time $t$, the total survival function for a transition by a particle in the compartment at time $t$ given that the last transition occurred at time $t'$ is expressed as
	\begin{equation}\label{eq_MarkovSurvival}
		\Phi_i(t|t')=\exp\left(-\int_{t'}^{t}\sum_{j=1}^{n_i}\mu_{i,j}(s)M_i(s)ds\right).
	\end{equation}
	We note that the transition rate, $\mu_{i,j}(t)$, can be more general and is not required to depend on the number of particles for birth and death processes. The survival function above is referred to as an inhomogeneous exponential distribution, and simulating from this distribution may be approached in a similar manner to a general inhomogeneous Poisson process. 
	
	For implementation, we refer readers to \cite{RossSheldonM2019Itpm}. Draws from this distribution can be found much more simply in the case where the rates are constant, and the distribution simplifies to an exponential distribution.
	
	For the non-Markovian compartments the probability of each particle transitioning will be dependent on the time of arrival of the particle. We are considering the case where the non-Markovian compartment is subject to both Markovian rates and a non-Markovian rate. 
	The evolution algorithm will need to store the time to the next non-Markovian transition for all particles. In our fractional compartmental model, the distribution of inter-event times is assumed to be Mittag-Leffler distributed with its survival function given by \eqref{eq_MLsurvival}. The waiting times may be sampled from this distribution as (see Appendix C),
	\begin{equation} \label{eq_DrawML}
		\Delta t=-\tau\ln(u)\left(\frac{\sin(\alpha \pi)}{\tan(\alpha \pi v)}-\cos(\alpha \pi)\right)^{\frac{1}{\alpha}},
	\end{equation}
	where $u$ and $v$ are two independent uniform $[0,\;1]$ random variables. We will assume that there are initially $M_i(0)$ particles that entered compartment $i$, more general initial conditions will be discussed later. With these considerations in place, the exact stochastic simulation method for a system of $N$ compartments is outlined as follows:
	
	\begin{enumerate}
		\item Initialize the system time, $t=0$.
		\item{For compartment $i$ with $n_i\geq 1$ associated Markovian transitions, generate $\Delta T_i$ from the survival function \eqref{eq_MarkovSurvival} using the method of choice, representing the combined time to the next transition for all particles in the compartment.}
		\item{If compartment $i$ also involves a non-Markovian removal process, generate the non-Markovian waiting times $\{\Delta t_{i,j}\}$ using \eqref{eq_DrawML} for $j\in\{1,2,\cdots,M_i(t)\}$.}
		\item{Determine the time to the next transition for the system as
			\[\Delta t=\min_{\forall i,j}(\Delta T_i,\{\Delta t_{i,j}\}).\]}
		\item{Either using the Gillespie algorithm or by $\argmin\{\Delta t_{i,j}\}$, determine the next transition process and update $M_i(t+\Delta t)$.}
		\item{Update the waiting times: $\{\Delta T_i\}\leftarrow\{\Delta T_i - \Delta t\}$ and $\{\Delta t_{i,j}\}\leftarrow\{\Delta t_{i,j}-\Delta t\}$.
			Redraw $\Delta T_i$ when it happens to be the next transition (when $\Delta T_i=0$) or if the corresponding survival function \eqref{eq_MarkovSurvival} has been modified following the transition. Track particles entering or leaving the non-Markovian compartment. For particles entering, generate a new non-Markovian waiting time and add it to $\{\Delta t_{i,j}\}$. For particles leaving, remove the corresponding waiting time from $\{\Delta t_{i,j}\}$.}
		\item{Update system time $t\leftarrow t+\Delta t$, and repeat from Step 4 onwards until some desired condition is met.}
	\end{enumerate}
	Due to the memoryless nature of the Markovian processes, arbitrary initial conditions should be only imposed on the non-Markovian compartments. Consider a scenario where the initial condition for a non-Markovian compartment $i$ is given by the arrival time list $\{t'_{i,j}\}$ for $j\in\{1,2,\cdots,M_i(0)\}$ and $t'_{i,j}<0$. This necessitates sampling from the conditional waiting time density for each arrival time
	\begin{equation}
		\psi_c(t|-t'_{i,j})=\frac{\psi(t-t'_{i,j})}{\Psi(-t'_{i,j})},
	\end{equation}
	which can be trivially done by rejection sampling:
	\begin{enumerate}
		\item[S1]{For particle $j$ with arrival time $t'_{i,j}$ in the non-Markovian compartment $i$, generate a waiting time $\Delta w$ using \eqref{eq_DrawML}.}
		\item[S2]{Accept the non-Markovian waiting time $\Delta t_{i,j}=\Delta w + t'_{i,j}$ if $\Delta w > -t_{i,j}$. Otherwise, go to step 1.}
		\item[S3]{Repeat Steps S2 and S3 for each particle $j$ in the compartment and for each non-Markovian compartment $i$.}
	\end{enumerate}
	This would replace Step 3 in the algorithm described in above.

	\section{Fractional Order Epidemiological Models}
	In the following sections, we aim to delineate the distinction between mean states of the system predicted by governing equations (see Appendix A \& B for details) and those obtained from the exact stochastic simulation, particularly evident when the system deviates from the continuum limit of large particle numbers. To illustrate this point, we will investigate two epidemiological models: the fractional order resusceptibility susceptible--infected--susceptible (SIS) model, and the fractional order susceptible--infected--recovered (SIR) model incorporating vital dynamics. The two models, along with their respective numerical simulation results, are presented in Sections \ref{Section_frSISModel} and \ref{Section_frSIRModel}. The two models, along with their respective numerical simulation results and subsequent discussions, are presented in Sections \ref{Section_frSISModel} and \ref{Section_frSIRModel}.
	\subsection{Fractional Order Resusceptibility SIS Model}\label{Section_frSISModel}
\begin{figure}[htbp]
		\begin{center}
			\includegraphics[width=0.7\linewidth]{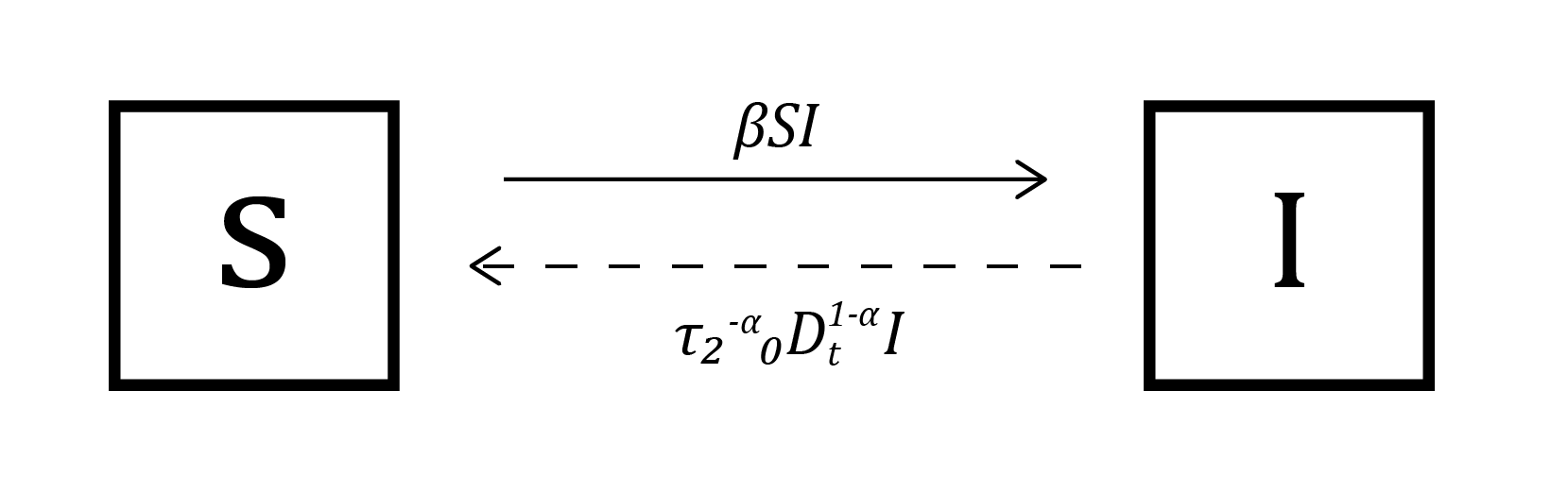}
			\caption{Flux flow of the fractional order resusceptibility SIS model.} 
			\label{fig_frSIS_Flux}
		\end{center}
		\vspace{0\baselineskip}
	\end{figure}

	The fractional resusceptibility SIS model is an extension of the standard SIS model. This model comprises a susceptible compartment, $S$, and an infectious compartment, $I$. The fractional order resusceptibility accounts for situations of chronic infection, wherein some individuals fail to recover from the disease. In addition, it assumes that individuals do not gain immunity to the disease, so they become susceptible again after recovery from the disease. In this model, infection follows a Markovian process, where susceptible individuals become infected at a transition rate determined by the law of mass action: $\beta SI$. Here the rate parameter, $\beta$, is typically considered as a constant, although a general time-dependent $\beta(t)$ could also be considered. On the other hand, recovery is a non-Markovian process, where infective individuals return to the susceptible compartment with a fractional order $0<\alpha\leq 1$ and a characteristic rate of $\tau_2^{-\alpha}$. The model contains no vital dynamics, making it a closed system with a fixed number of individuals, $N=S(t)+I(t)$. The schematic of the fractional SIS model is shown in Figure \ref{fig_frSIS_Flux}.
	\vspace{-1\baselineskip}
	
	To construct the governing equations for this fractional order two-compartment model, we start with \eqref{eq_GM2} and let $u_1=S$ and $u_2=I$. For simplicity of the example, we consider the case where the initial populations of the infective individuals ($i_0$) and susceptible individuals ($s_0$) are injected into the corresponding compartments at $t=0$. The arrival flux into the infectious compartment originates from the Markovian removal flux and the initial injection, thus $q_2(t)=s_0\delta(t-0^+)+\beta SI$ and $q^+_2(t)=\beta SI$. The infectious compartment contains no Markovian removal process so $\omega_2(t)=0$ and $\Theta_i(t, 0)=1$. The governing equations for the model can be derived through a flux balance consideration:
	\begin{equation} \label{eq_fSIS1}
		\begin{split}
			\frac{dS}{dt}&=-\beta SI+{\tau_2}^{-\alpha}{}_{0} \mathcal{D}_t^{1-\alpha}I,\\
			\frac{dI}{dt}&=\beta SI-{\tau_2}^{-\alpha}{}_{0}\mathcal{D}_t^{1-\alpha}I.\\
		\end{split}
	\end{equation}
	
	We consider the fractional resusceptibility SIS model with parameters: $\alpha=0.95$, $\tau_2=1$, and $\beta=2/N_0$, where $N_0 = s_0 + i_0$ denotes the initial total population. The inverse proportionality of $\beta$ to $N_0$ suggests that as the initial total population $N_0$ increases, the rate at which susceptible individuals encounter infected individuals becomes constrained. For initial conditions, we consider two scenarios: (1) Small population dynamics with $s_0=98$ and $i_0=2$ at $t=0$; (2) Increased initial injections scaled up by a factor of five, i.e., $s_0=490$ and $i_0=10$ at $t=0$, where extinction events are less likely to occur. With the model having only two compartments and a fixed total population, its complete dynamics can be inferred from either compartment. Here, we focus on showcasing the dynamics in the infectious compartment.
	\begin{figure}[!htbp]
		\begin{center}
			\begin{subfigure}{.49\textwidth} 
				\includegraphics[width=1\linewidth]{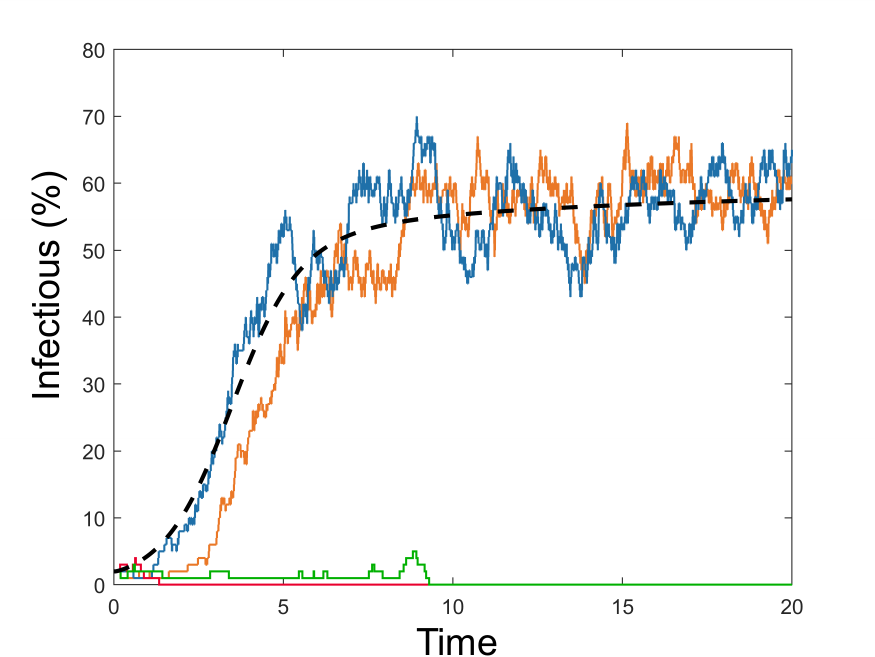}
				\caption{}	
				\label{fig4:sub1}
			\end{subfigure}
			\begin{subfigure}{.49\textwidth}
				\centering
				\includegraphics[width=1\linewidth]{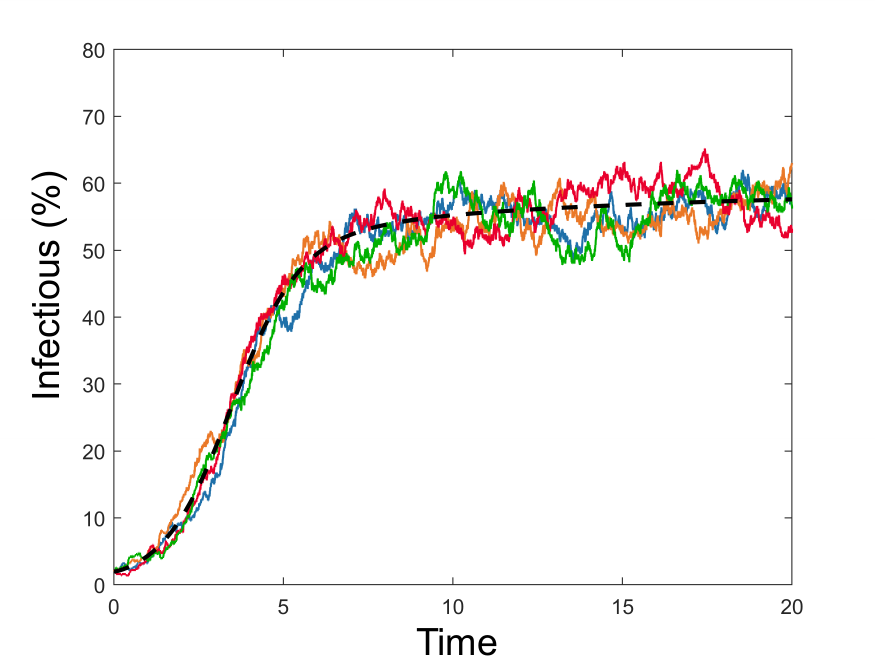}
				\caption{}
				\label{fig4:sub2}
			\end{subfigure}
			\caption{Representative sample paths showing the temporal dynamics of the infectious population for the stochastic fractional resusceptibility SIS model with $\alpha=0.95$, alongside the deterministic solution depicted by the dashed black line. (a) Sample paths start with initial conditions $s_0=98$ and $i_0=2$ (total initial population $N_0=100$). The red and green sample paths represent two early extinction events of the disease. (b) Sample paths starts with scaled initial conditions $s_0=490$ and $i_0=10$ ($N_0=500$). The other parameters are $\tau_2=1$ and $\beta=2/N_0$. The deterministic solution was solved using the DTRW method with $\Delta t=0.05$. } 
			\label{fig_SPs}
			\vspace{-1.5\baselineskip}
		\end{center}
	\end{figure}
	The model can also be described within the DTRW framework, offering an efficient numerical method for approximating the dynamics of the mean states of the governing equations. This is accomplished by utilizing \eqref{eq_DTRWME1}, \eqref{eq_DTRWME2}, and \eqref{eq_SibuyaKernel}. When a specific $\Delta{t}$ is chosen, the parameters in the DTRW formulation are linked to the model parameters in the following manner: $\omega(n)=1-\exp({-\beta\Delta t})$ and $r = (\Delta{t}/\tau_2)^\alpha$. The initial conditions are implemented by taking $S(0)=s_0$ and $I(0)=i_0$ for the corresponding scenarios. 
	
	In Figure \ref{fig4:sub1}, we present four sample paths illustrating the dynamics of a small population. The dashed black line represents the mean state of the infectious compartment based on the governing equations, as computed using the DTRW method. The red and green lines depict scenarios where an early extinction event occurs. Notably, the volatility observed in small population dynamics, as depicted by the blue and orange sample paths, is reduced when the initial population in each compartment is scaled up by a factor of five, as demonstrated in Figure \ref{fig4:sub2}.
	
	\begin{figure}[!htbp]
		\centering
		\begin{subfigure}{.5\textwidth}
			\centering
			\includegraphics[width=1\linewidth]{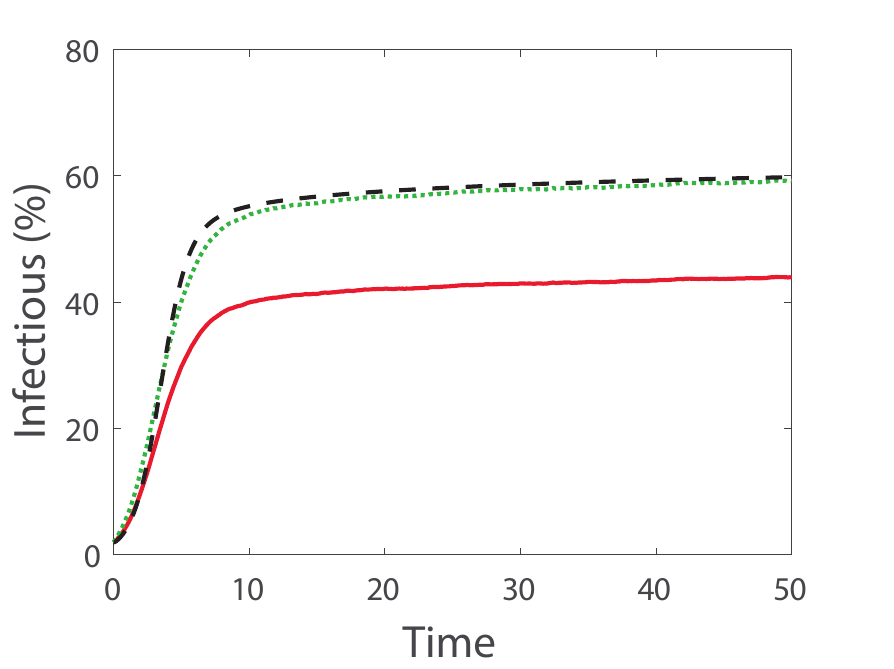}
			\subcaption{}
			\label{fig:sub1}
		\end{subfigure}%
		\begin{subfigure}{.5\textwidth}
			\centering
			\includegraphics[width=1\linewidth]{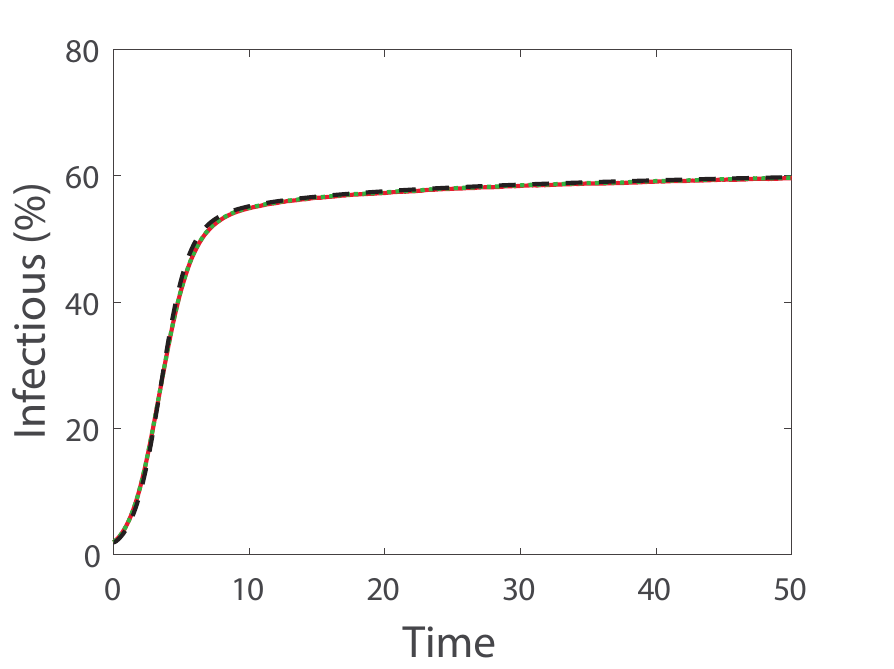}
			\subcaption{}
			\label{fig:sub2}
		\end{subfigure}
		\begin{subfigure}{.49\textwidth}
			\centering
			\includegraphics[width=1\linewidth]{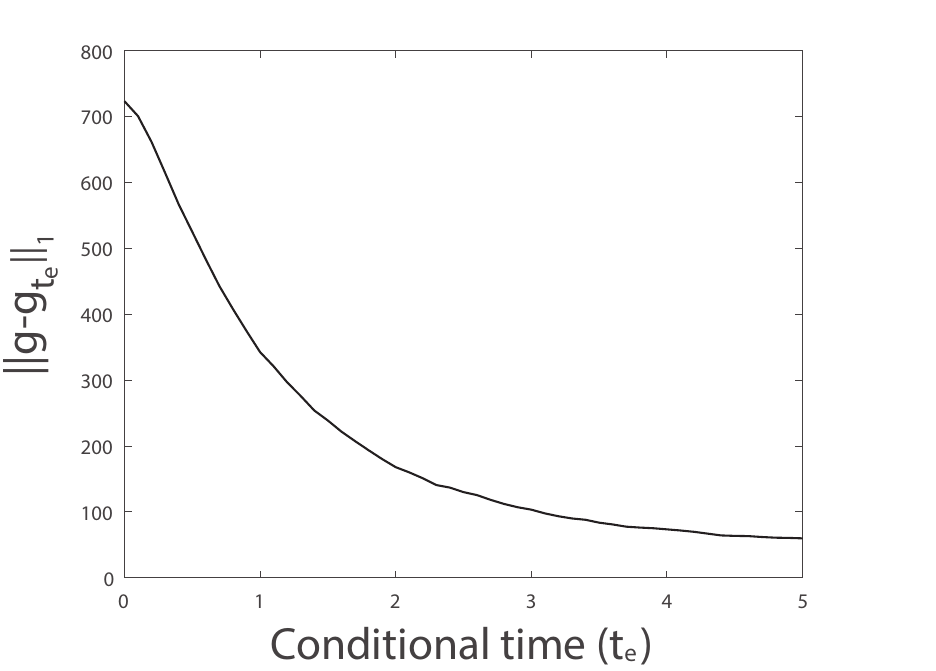}
			\subcaption{}
			\label{fig:sub3}
		\end{subfigure}
		\begin{subfigure}{.49\textwidth}
			\centering
			\includegraphics[width=1\linewidth]{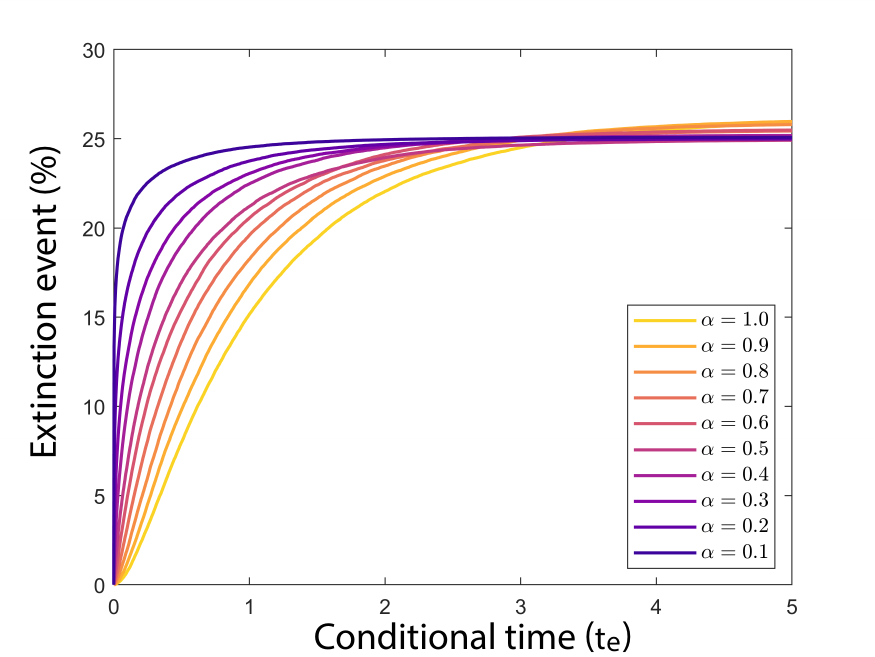}
			\subcaption{}
			\label{fig:sub4}
		\end{subfigure}%
		\caption{Dynamics of the mean state of the infectious compartment in the fractional resusceptibility SIS model with $\alpha=0.95$. (a) Colored lines show the mean percentage of infectious population versus time. The red line represents the overall mean, while the green dotted line reflects the mean conditioned on no early extinction for $t\in\left[0,\;50\right]$. The dashed black line represents the deterministic solution calculated using the DTRW numerical scheme with $\Delta t = 0.05$. The stochastic simulation results are obtained with initial injections $s_0=98$ and $i_0=2$ ($N=100$). (b) Dynamics of the mean state with scaled populations ($s_0=490$, $i_0=10$, and $N=500$). (c) The error measure, $||g-g_{t_{e}}||_{1}$ as a function of $t_{e}$ in the case of small population dynamics. (d) The percentage of paths conditioned on occurrence of the extinction event prior to time $t_{e}$ as a function of $t_{e}\in[0,50]$, the coloured lines showed behavior for different values of $\alpha \in (0,1]$. Note that (c) and (d) are concerned with the case of the small population dynamics. The parameters for the stochastic simulations are $\beta=0.02$, and $\tau_2=1$. Each mean value is calculated by averaging over $10000$ sample paths for (a)-(c), and $100000$ sample paths for (d).}
		\label{fig:test}
		\vspace{-1.5\baselineskip}
	\end{figure}
	
	The strong deviation of the mean state dynamics of the stochastic process from the governing equations is evident in Figure \ref{fig:sub1} for small populations. The mean state of the stochastic process, depicted in red, significantly differs from the mean state of the governing equations represented by the dashed black line. This discrepancy primarily arises from early extinction events of the disease, as evidenced by the observation that the mean state dynamics conditioned on no early extinction in the time interval $t\in\left[0,\;50\right]$ exhibit a closer alignment with the mean state of the governing equations.  In contrast, Figure \ref{fig:sub2} demonstrates the case of a larger initial population, where the mean behavior of the stochastic process is more accurately captured by the governing equations. The improved alignment is a result of the decreased likelihood of extinction events. To further examine the impact of extinction events on the mean state, we let $g$ denote the mean state obtained using the DTRW method, and $g_{t_{e}}$ denote the mean state obtained from the stochastic simulation, conditioned on no extinction events prior to time $t_{e}$. The convergence between the two solutions is demonstrated by plotting the $L_1$ norm of the differences between $g$ and $g_{t_{e}}$, $||g-g_{t_{e}}||_{1}$, as a function of $t_{e}$, as shown in Figure \ref{fig:sub3}. Here, the $L_1$ norm of a function $f(t)$ is defined as
	\begin{equation}
		||f||_{1}=\int_{0}^{t_{s}}|f(t')|dt',
	\end{equation}
	where $t_{s}$ represents the simulation time, which is set to $50$ in this example.\\
	
	To investigate the impact of the fractional order dynamics on the probability of early extinction, we plot the percentage of paths that go extinct before time $t_{e}$ as a function of $t_{e}$ for different values of $\alpha$, as shown in Figure \ref{fig:sub4}. The plot reveals that the majority of extinction events occur prior to $t=5$ and plateaus at a percentage around $25\%$ for all values of $\alpha$. Interestingly, we observe that the probability of early extinction plateaus more quickly for lower values of $\alpha$. This implies that the behavior of the mean state, as obtained from stochastic simulations, diverges more significantly from that predicted by the governing equations in the case of fractional order dynamics, especially at short times, when compared to integer order dynamics. This finding highlights the critical role of stochastic simulation in accurately capturing the dynamics of fractional order compartment models.
	
	\subsection{Fractional Order Recovery SIR Model with Vital Dynamics}\label{Section_frSIRModel}
\begin{figure}[htbp]
		\begin{center}
			\includegraphics[width=0.9\linewidth]{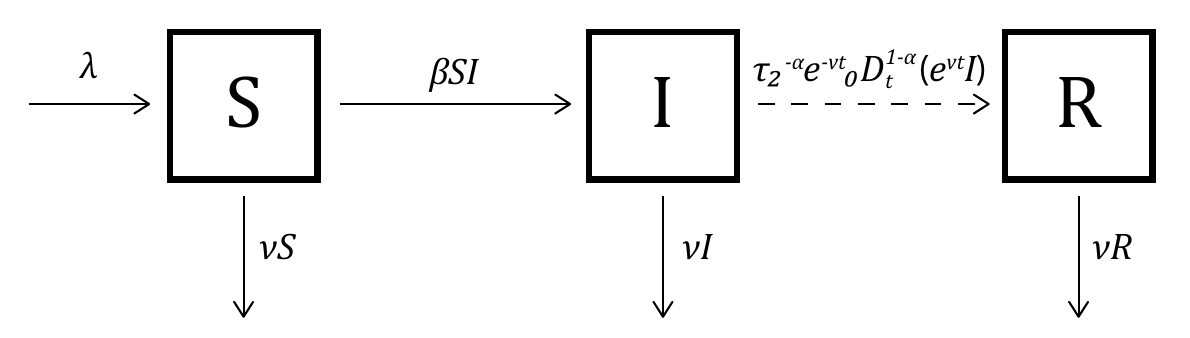}
			\caption{Flux flow of the fractional order recovery SIR model.} 
			\label{fig_frSIR_Flux}
		\end{center}
		\vspace{-1\baselineskip}
	\end{figure}

	Similar to the fractional order resusceptibility SIS model, the fractional order recovery model extends the standard SIR model to incorporate the effects of chronic infection. This model divides the population into three compartments: susceptible $S$, infectious $I$, and recovered $R$. Unlike the SIS model, individuals who recover from the disease transition to the recovered compartment $R$, indicating that they have gained immunity to the disease.
	Once again, the transition from susceptible $S$ to infectious $I$ is modeled as a Markovian process with a transition rate of $\beta SI$. Recovery from the infectious compartment $I$ to the recovered compartment $R$ is characterized by a fractional order process with an exponent $0<\alpha\leq 1$ and characteristic rate $\tau_2^{-\alpha}$. Furthermore, the model incorporates vital dynamics, including a constant birth rate $\lambda$ into the susceptible compartment $S$ and population-dependent death rates from each compartment with a rate parameter $\nu$, as illustrated in Figure \ref{fig_frSIR_Flux}.\\
	
	Consider the case where the initial conditions are given by the injections of infective ($i_0$), susceptible ($s_0$) and recovered ($r_0$) individuals at $t=0$. Formulating the governing equations akin to those outlined in Section \ref{Section_frSISModel}, we arrive at a system of fractional order differential equations:
	\begin{equation}
		\begin{split}
			\frac{dS}{dt}&=\lambda -\beta SI -\nu S,\\
			\frac{dI}{dt}&=\beta SI  - {\tau_2}^{-\alpha}\exp(-\nu t){}_{0} \mathcal{D}_t^{1-\alpha}\left(\exp(\nu t)I\right) - \nu I,\\
			\frac{dR}{dt}&= {\tau_2}^{-\alpha}\exp(-\nu t){}_{0} \mathcal{D}_t^{1-\alpha}\left(\exp(\nu t)I\right) - \nu R.
		\end{split}
	\end{equation}
	
	We examine the fractional recovery SIR model, defined by the governing equations above, across a range of $\alpha$ values from $0.3$ to $0.9$, alongside other parameters $\tau_2=1$, and $\beta=2/N_0$, where $N_0 = s_0 + i_0 + r_0$. Again, two initial condition scenarios are considered: (1) Small population dynamics with $s_0=95$ and $i_0=5$ at $t=0$; (2) Larger initial injections, scaled by a factor of ten, i.e., $s_0=950$ and $i_0=50$. The model parameters are linked to those of the DTRW numerical scheme as follows: $\Lambda = \lambda\Delta t$, $\omega(n)=1-\exp(-\beta\Delta t)$, $\zeta(n) = 1-\exp(-\nu\Delta t)$, and $r = (\Delta{t}/\tau_2)^\alpha$. The initial conditions are set as $S(0)=s_0$, $I(0)=i_0$, and $R(0) = r_0$ for the respective scenarios.
	
	\begin{figure}[!htbp]
		\centering
		\begin{subfigure}{.5\textwidth}
			\centering
			\includegraphics[width=1\linewidth]{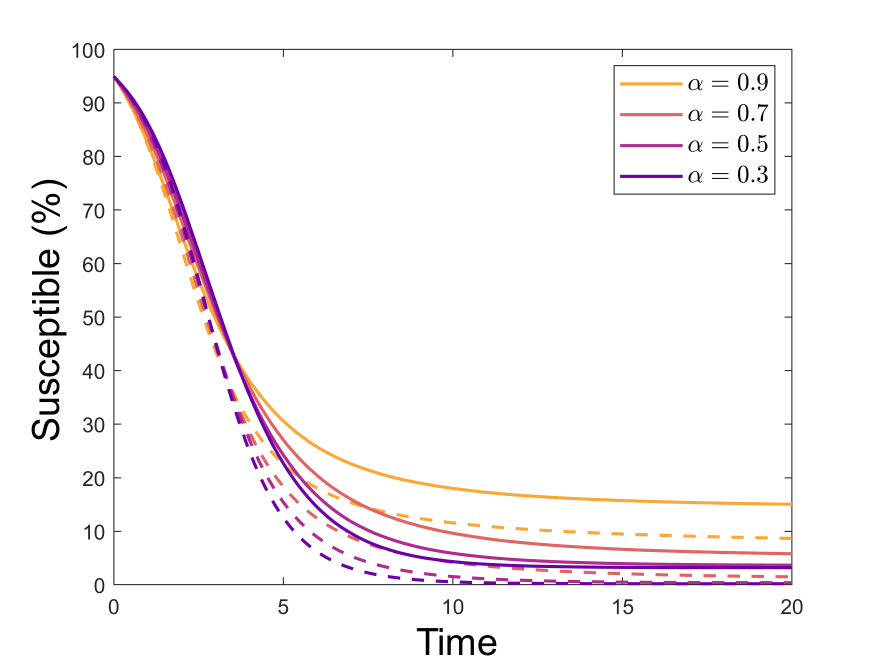}
			\caption{}
			\label{fig6:sub1}
		\end{subfigure}%
		\begin{subfigure}{.5\textwidth}
			\centering
			\includegraphics[width=1\linewidth]{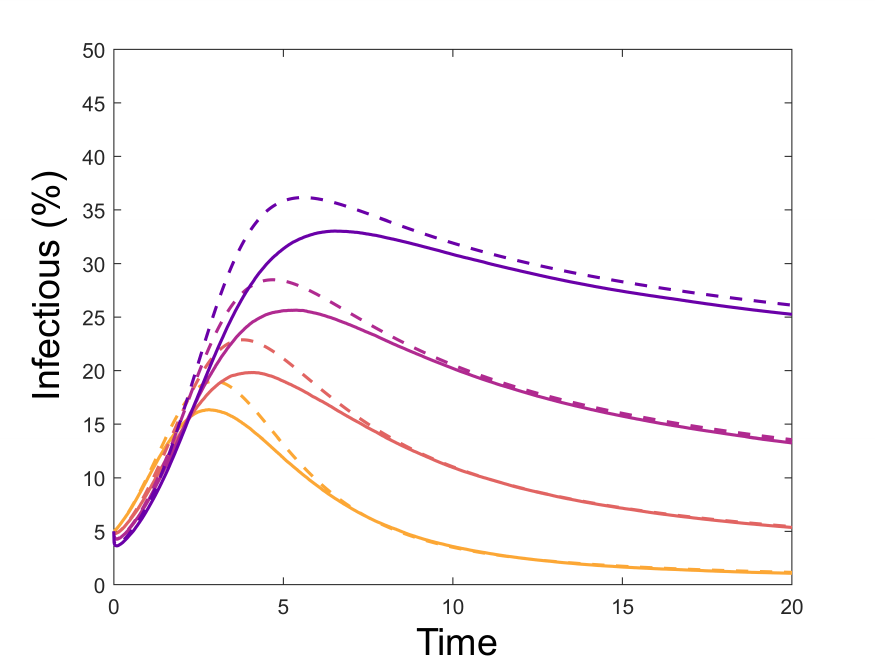}
			\caption{}
			\label{fig6:sub2}
		\end{subfigure}
		\begin{subfigure}{.49\textwidth}
			\centering
			\includegraphics[width=1\linewidth]{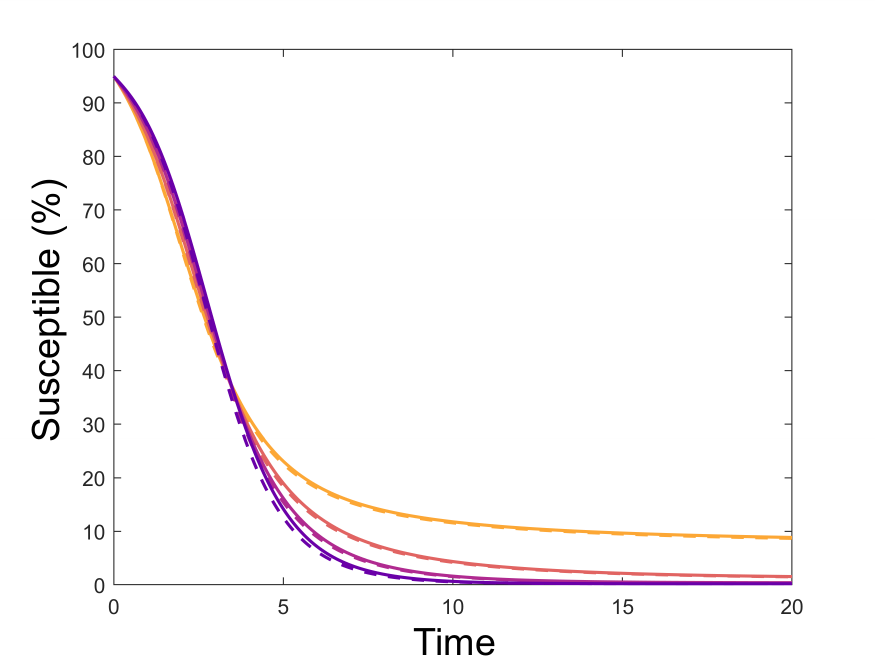}
			\caption{}
			\label{fig6:sub3}
		\end{subfigure}
		\begin{subfigure}{.49\textwidth}
			\centering
			\includegraphics[width=1\linewidth]{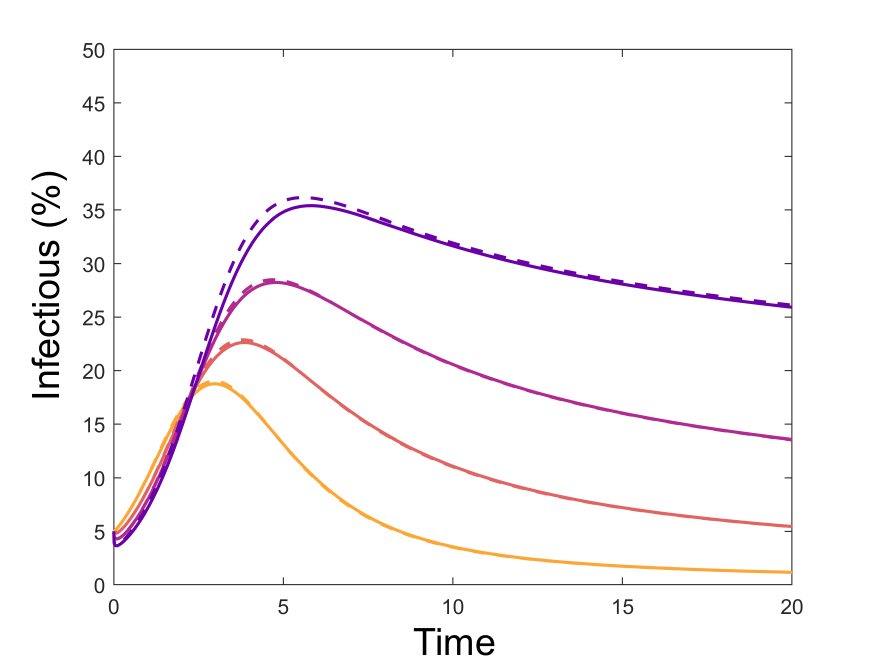}
			\caption{}
			\label{fig6:sub4}
		\end{subfigure}%
		\vspace{1\baselineskip}
		\caption{Impact of population size on the mean state of susceptible and infectious compartments in the fractional recovery SIR model with $\alpha= 0.3, 0.5, 0.7, 0.9$. The color gradient from dark to light indicates the increasing $\alpha$ values. Panels (a) and (b) depict the temporal dynamics of susceptible and infectious populations, initialized with $s_0=95$, $i_0=5$, and $r_0 = 0$ (total initial population $N_0=100$). Solid lines represent the stochastic solution of the mean state, while the dashed lines depict the deterministic solution. Panels (c) and (d) display the corresponding plots for large population dynamics with scaled initial injections $s_0=950$, $i_0=50$, and $r_0 = 0$ ($N_0=1000$). Other model parameters include $\lambda = N_0/1000$, $\beta = 2/N_0$, $\nu = 0.001$, $\tau_2=1$. Each stochastic solution was averaged over 10000 sample paths, while the deterministic solution was solved using the DTRW numerical scheme with $\Delta t = 0.00002.$}
		\label{fig6}
		\vspace{-1.5\baselineskip}
	\end{figure}
	
	The impact of population size on the mean state dynamics across various $\alpha$ values is illustrated in Figure \ref{fig6}. Substantial deviations between the mean state dynamics of the stochastic process and the governing equations are noticeable in Figures \ref{fig6:sub1} and \ref{fig6:sub2} for smaller populations. However, increasing the population size by a factor of ten significantly improves the alignment between stochastic and deterministic solutions, as demonstrated in Figures \ref{fig6:sub3} and \ref{fig6:sub4}. These observations persist across all $\alpha$ values considered.
	
	In contrast to the fractional resusceptibility SIS model, where the primary issue causing discrepancy between stochastic and deterministic solutions is early extinction, the fractional recovery SIR model also experiences challenges with the long-term dynamic behavior of the infectious population. After the initial relaxation time, the fraction of the infectious population becomes exceedingly small, particularly for $\alpha$ values close to one. In such cases, the system almost necessitates an infinite population size for solutions to converge. This scenario can be analogized to the 'atto-fox' problem in predator-prey models.
	
	\begin{figure}[!htbp]
		\centering
		\begin{subfigure}{.5\textwidth}
			\centering
			\includegraphics[width=1\linewidth]{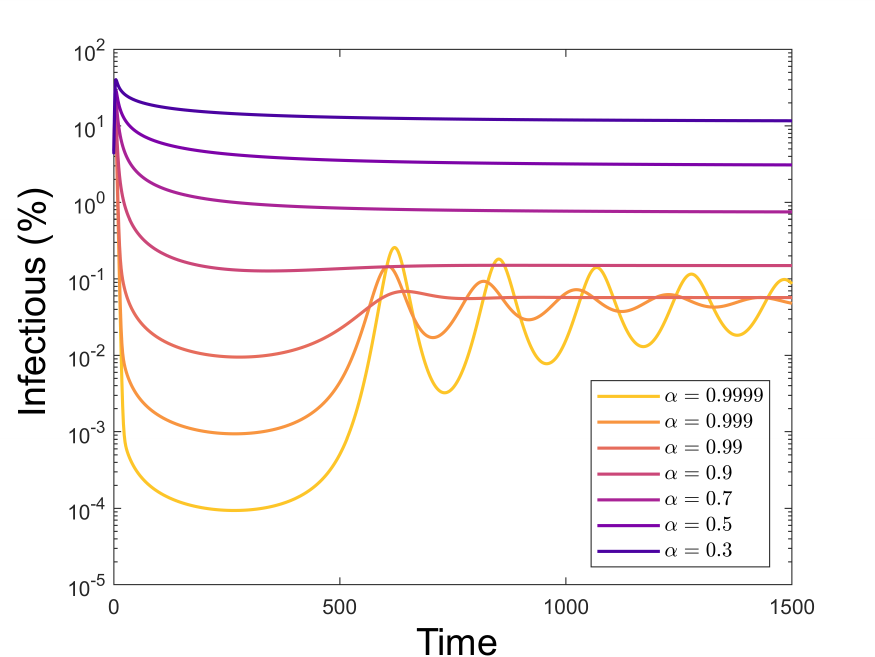}
			\caption{}
			\label{fig7:sub1}
		\end{subfigure}%
		\begin{subfigure}{.5\textwidth}
			\centering
			\includegraphics[width=1\linewidth]{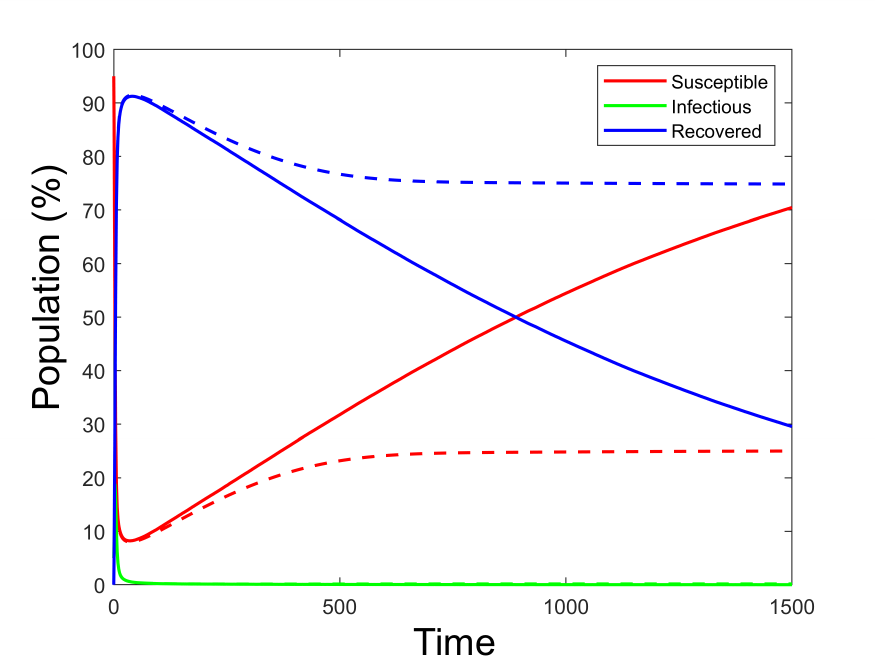}
			\caption{}
			\label{fig7:sub2}
		\end{subfigure}
		\caption{Divergence of stochastic and deterministic mean state dynamics at long times in the fractional recovery SIR model. (a) Time evolution of infectious populations as a percentage in the deterministic model with various $\alpha$ values ($\alpha = 0.3, 0.5, 0.7, 0.9, 0.99, 0.999, 0.9999$). The color gradient from dark to light indicates the increasing $\alpha$ values. (b) The mean state dynamics between stochastic and deterministic models begin to diverge over prolonged periods with $\alpha = 0.9$, despite their alignment during the initial relation time (see Figures \ref{fig6:sub3} and \ref{fig6:sub4}). Other model parameters include $\lambda = N_0/1000$, $\beta = 2/N_0$, $\nu = 0.001$, $\tau_2=1$, with initial conditions $s_0=950$, $i_0=50$, and $r_0 = 0$ ($N_0=1000$). Each stochastic solution was averaged over 10000 sample paths, while the deterministic solution was solved using the DTRW numerical scheme with $\Delta t = 0.01.$}
		\label{fig7} 
		\vspace{-1.5\baselineskip}
	\end{figure}
	
	Even when considering relatively low $\alpha$ values, the long-term endemic steady state of the infectious population diminishes to such an extent that it becomes prone to stochastic fluctuations and risks extinction, as illustrated in Figure \ref{fig7:sub1}. This vulnerability is further underscored in Figure \ref{fig7:sub2}, where the long-term stochastic solution of the mean state in the context of a large population, with $\alpha = 0.9$, starts to deviate from the deterministic solution after approximately $t=200$. This deviation occurs despite their initial alignment during the relaxation phase, as observed in Figures \ref{fig6:sub3} and \ref{fig6:sub4}.
	
	\section{Conclusion}
	
	In this study, we have focused on fractional order compartment models, which originate from an underlying stochastic process governing particle dynamics, thereby ensuring mass conservation \cite{AngstmannFOCM2017}. Within this framework, each particle in the non-Markovian compartment will undergo a semi-Markov process, where inter-event times within compartments follow the Mittag-Leffler waiting time distributions, giving rise to Riemann-Liouville fractional derivatives in the governing equations. To address these models, we have introduced an exact stochastic simulation algorithm. Our algorithm tracks the waiting time for the non-Markovian removal process for each particle, consolidating Markovian processes' waiting times in each compartment using the Gillespie algorithm. Then, the system evolves based on the minimum of these stored waiting times, akin to the Next Reaction Method.
	
	Our work underscores the significance of inherent stochasticity, exemplified through epidemiological models including a fractional resusceptibility SIS model and a fractional recovery SIR model. We demonstrated that, as the total population increases, the mean state of the governing equation aligns with stochastic simulation results. However, in cases of small populations, early extinction events in the stochastic simulation lead to divergent solutions. We further investigated the impact of fractional dynamics on early extinctions. In addition, stochastic fluctuations can impact the behavior of a dynamical system not only in its early stages but also over extended periods, depending on dynamics of the system. Specifically, if the steady state of a dynamical system tends toward zero at long times, it can be significantly influenced by stochastic fluctuations through nonlinear responses, as demonstrated in the fractional recovery SIR model. We note that convergence between mean state ontained from stochastic simulation and governing equations will be assured regardless of the population size, as long as the arrival flux $q_i(t)$ is deterministic or linearly dependent on the compartment's state due to the linearity of \eqref{eq_U2}. While we have numerically demonstrated the convergence in the case of a nonlinear flux within the investigated epidemiological models, the mathematical proof would require model-specific analysis akin to the approach demonstrated for the integer order case by Armbruster et al. \cite{Armbruster2017a,Armbruster2017b}.
	
	In conclusion, our research highlights the importance of embracing the inherent stochasticity in modelling when dealing with complex systems with a small population, and our algorithm offers a precise and adaptable means to achieve this. By enabling exploration of diverse fractional order compartment models, it fosters a deeper understanding of real-world phenomena governed by fractional order dynamics. Moreover, the fractional order compartment model simulation, facilitated by the algorithm developed herein, lays the groundwork for implementing Approximate Bayesian Computation (ABC) in parameter estimation \cite{Toni2008}. It is worth noting that inference for fractional order models remains an active research area, presenting challenges, particularly in dealing with finite time windows and the limited sample size of experimental data \cite{Ralf2018,Zhang2023,Chahoy2010}.
	
	\section*{Appendix A: Discrete Time Fractional Order Resusceptibility SIS Model}\label{DTRW_frSIS}
	The use of DTRWs as a basis of a numerical method for solving fractional order partial differential equations, and fractional order compartment models, was introduced in \cite{AngstmannDTRW2015}
	and \cite{AHJM2019}, respectively. Here, we derive the DTRW numerical scheme for the fractional resusceptibility SIS model described in Section \ref{Section_frSISModel}. 
	
	To model recovery as a fractional order removal process, the probability of an infected individual re-entering the susceptible compartment will depend on the number of time steps since it arrived at the infectious compartment. To ensure the discrete time process limits to the continuum process as time step, $\Delta t$, goes to zero, it is necessary to introduce an additional self-jump process so that the individuals in the infectious compartment have a certain probability, $r$, to enter the subsceptible compartment, conditioned on that the non-Markovian transition occurs. Furthermore, it is assumed that if the individual was infected before the first time step, they will just transition into the susceptible compartment with probability 1. Let $\omega(n)$ denote the probability of an infected individual comming in contact and infecting a susceptible in the $n^{\text{th}}$ time step, and $\gamma(n)$ denote the probability of making a non-Markovian transition on the $n^{\text{th}}$ step, conditioned on surviving the transition for the first $n-1$ steps. In these notations, the probability of surviving the non-Markovian transition for $n$ steps is given by
	\begin{equation}
		\Psi(n) = \prod_{j=0}^n\left(1-\gamma(j)\right),
	\end{equation}
	with $\gamma(0)=0$. From this, the arrival flux entering the infectious compartment can be iteratively written as
	\begin{equation}\label{eq_DTRW1}
		Q_{I}^{+}(n) = \sum_{k=-\infty}^{n-1}\omega(n)S(n-1)\Psi(n-k-1)Q_{I}^{+}(k)+(1-r)\sum_{k=0}^{n-1}\gamma(n-k)\Psi(n-k-1)Q_{I}^{+}(k).
	\end{equation}
	Since we have assumed that initial injections happens at time zero (i.e., at $n=0$), we have
	\begin{equation}
		Q_{I}^{+}(k) = i_0\delta_{k,0} \text{\quad ($k \leq 0$),}
	\end{equation}
	here $\delta_{k,0}$ is the Kronecker delta function. Thus \eqref{eq_DTRW1} may be expressed as
	\begin{equation}\label{eq_DTRW2}
		\begin{split}
			Q_{I}^{+}(n) =& \sum_{k=0}^{n-1}\omega(n)S(n-1)\Psi(n-k-1)Q_{I}^{+}(k)\\
			& +(1-r)\sum_{k=0}^{n-1}\gamma(n-k)\Psi(n-k-1)Q_{I}^{+}(k).
		\end{split}
	\end{equation}
	The number of infectious individuals on the $n^{\text{th}}$ step is
	\begin{equation}\label{eq_DTRW3}
		I(n) = \sum_{k=0}^{n} \Psi(n-k)Q_{I}^{+}(k).
	\end{equation}
	The increment for the infectious individuals is 
	\begin{equation}\label{eq_DTRW4}
		\begin{split}
			I(n)-I(n-1) &= Q_{I}^{+}(n) + \sum_{k=0}^{n-1}(\Psi(n-k)-\Psi(n-k-1))Q_{I}^{+}(k)\\
			&= Q_{I}^{+}(n) - \sum_{k=0}^{n-1}\gamma(n-k)\Psi(n-k-1)Q_{I}^{+}(k) .\\
		\end{split}
	\end{equation}
	Substituting \eqref{eq_DTRW2} into the right-hand side of \eqref{eq_DTRW4} and using \eqref{eq_DTRW3} gives
	\begin{equation}\label{eq_DTRW5}
		\begin{split}
			I(n)-I(n-1) &= \omega(n)S(n-1)I(n-1)-r\sum_{k=0}^{n-1}\gamma(n-k)\Psi(n-k-1)Q_{I}^{+}(k).
		\end{split}
	\end{equation}
	To proceed further, we employ the unilateral Z-transform, which transforms a function $F(n)$ from $n$ to $z$ domain as
	\begin{equation}
		\mathcal{Z}\{F(n)|z\}=\sum_{n=0}^{\infty}F(n)z^{-n}.
	\end{equation}
	Applying the Z-transform to Eq.(\ref{eq_DTRW3}) and using its convolution property we have
	\begin{equation}
		\mathcal{Z}\{I(n)|z\}=\mathcal{Z}\{\Psi(n)|z\}\mathcal{Z}\{Q_I^+(n)|z\}.
	\end{equation}
	Likewise,
	\begin{equation}
		\mathcal{Z}\{\sum_{k=0}^{n-1}\gamma(n-k)\Psi(n-k-1)Q_{I}^{+}(k)|z\}=\mathcal{Z}\{\gamma(n)\Psi(n-1)|z\}\mathcal{Z}\{Q_I^+(n)|z\}.
	\end{equation}
	This enables Eq.(\ref{eq_DTRW5}) to be written in the form
	\begin{equation}\label{eq_DTRWME1}
		\begin{split}
			I(n)-I(n-1) &= \omega(n)S(n-1)I(n-1)
			-r\sum_{k=0}^{n}\kappa(n-k)I(k),
		\end{split}
	\end{equation}
	where $\kappa(n)$ is the discrete memory kernel defined as
	\begin{equation}\label{eq_DTRWkernel}
		\mathcal{Z}\{\kappa(n)|z\}=\frac{\mathcal{Z}\{\gamma(n)\Psi(n-1)|z\}}{\mathcal{Z}\{\Psi(n)|z\}}.
	\end{equation}
	Again, a flux balance consideration gives the governing equation for the susceptible compartment,
	\begin{equation}\label{eq_DTRWME2}
		S(n)-S(n-1) = -\omega(n)S(n-1)I(n-1)
		+r\sum_{k=0}^{n}\kappa(n-k)I(k).
	\end{equation}
	The DTRWs \eqref{eq_DTRWME1} and \eqref{eq_DTRWME2} converge to the governing equations of the fractional resusceptibility SIS model, \eqref{eq_fSIS1}, as $\Delta t \rightarrow 0$ and $r \ \rightarrow 0$, with
	\begin{equation}
		\lim_{\Delta t \rightarrow 0, r \rightarrow 0}\frac{r}{\Delta t^{\alpha}}=\tau^{-\alpha},
	\end{equation}
	provided that the probability mass function of the waiting steps follows the Sibuya($\alpha$) distribution, characterized by the survival function:
	\begin{equation}
		\Psi(n) = \frac{\Gamma(n-\alpha+1)}{\Gamma(n+1)\Gamma(1-\alpha)}
	\end{equation}
	where the probability of making a trnasition on the $n^{\text{th}}$ step, conditioned on surviving the first $n-1$ steps is given by
	\begin{equation}
		\gamma(n)=
		\begin{cases}
			0 & n=0,\\
			\frac{\alpha}{n} & n\geq1.
		\end{cases}
	\end{equation}
	The memory kernel can be obtained by applying the inversion of the Laplace transform to Eq. (\ref{eq_DTRWkernel}), yielding
	\begin{equation}\label{eq_SibuyaKernel}
		\kappa(n)= \frac{\Gamma(n+\alpha-1)}{\Gamma(n+1)\Gamma(\alpha-1)}-\delta_{0,n}+\delta_{1,n}.
	\end{equation}
	We see that $\kappa(0)=0$, $\kappa(1)=\alpha$ and $\kappa(2)=\frac{\alpha}{2}(\alpha-1)$. By exploiting a recursion relation, the memory kernel for $n\geq3$ can be efficiently computed, 
	\begin{equation}\label{eq_SibuyaKernel2}
		\kappa(n)= \left(1+\frac{\alpha-2}{n}\right)\kappa(n-1).
	\end{equation}
	The continuous time fractional order resusceptibility SIS model can be numerically solved with the DTRWs model using \eqref{eq_DTRWME1}, \eqref{eq_DTRWME2}, and \eqref{eq_DTRWkernel}.
	\section*{Appendix B: Discrete Time Fractional Order Recovery SIR Model}\label{DTRW_frSIR}
	To construct the DTRW numerical scheme for the fractional recovery SIR model described in Section \ref{Section_frSIRModel}, the following parameters are defined: $\Lambda$ the constant birth rate into the susceptible compartment S; $r$ the probability of an infected entering the recovered compartment $R$ given that the non-Markovian transition occurs; $\omega(n)$ the probability of an infected individual comming in contact with a susceptible resulting in an infection event in the $n^{\text{th}}$ time step; $\zeta(n)$ the probability that an individual will die on the $n^{\text{th}}$ step; $\gamma(n)$ the probability of making a non-Markovian transition on the $n^{\text{th}}$ step, conditioned on surviving the transition for the first $n-1$ steps. Again we assume that initial injection of infected individuals happens at $n=0$.
	
	The probability of an infected individual surviving the non-Markovian transition for $n$ steps is geven by
	\begin{equation}
		\Psi(n) = \prod_{j=0}^n\left(1-\gamma(j)\right),
	\end{equation}
	and the probability of an individual who entered the infectious compartment $I$ on the $k^{\text{th}}$ step has survived the death process until the $n^{\text{th}}$ step is
	\begin{equation}
		\Theta(n,k) = \prod_{j=k}^n\left(1-\zeta(j)\right).
	\end{equation}
	It follows that the arrival flux entering the infectious compartment is simply
	\begin{equation}
		\begin{split}
			Q_{I}^{+}(n) =& \sum_{k=0}^{n-1}\omega(n)S(n-1)\Theta(n-1,k)\Psi(n-k-1)Q_{I}^{+}(k)\\
			& +(1-r)\sum_{k=0}^{n-1}\gamma(n-k)\Theta(n,k)\Psi(n-k-1)Q_{I}^{+}(k),
		\end{split}
	\end{equation}
	and 
	\begin{equation}
		I(n) = \sum_{k=0}^{n} \Theta(n,k)\Psi(n-k)Q_{I}^{+}(k).
	\end{equation}
	Following a similar procedure as outlined in \eqref{eq_DTRW4}--\eqref{eq_SibuyaKernel2}, employing the Z-transform method, while leveraging the semi-group property of $\Theta(n,k)$,
	\begin{equation}
		\Theta(n,0) = \Theta(n,k) \Theta(k,0),
	\end{equation}
	the discrete time master equations for the fractional recovery SIR model is given by
	\begin{equation}
		\begin{split}
			S(n)-S(n-1) &= \Lambda -\omega(n)S(n-1)I(n-1) - \zeta(n)S(n-1),\\
			I(n)-I(n-1) &= \omega(n)S(n-1)I(n-1)  - \zeta(n)I(n-1) - r\Theta(n,0)\sum_{k=0}^{n}\kappa(n-k)\frac{I(k)}{\Theta(k,0)},\\
			R(n)-R(n-1) &= -\zeta(n)R(n-1) + r\Theta(n,0)\sum_{k=0}^{n}\kappa(n-k)\frac{I(k)}{\Theta(k,0)},
		\end{split}
	\end{equation}
	where the memory kernel $\kappa(n)$ is define as \eqref{eq_SibuyaKernel}.
	
	\section*{Appendix B: Sampling Mittag-Leffler Waiting Times}
	The survival function of Mittag-Leffler waiting time distribution can be expressed in terms of the two-parameter Mittag-Leffler function as,
	\begin{equation}
		\Psi(t) = E_{\alpha,1}\left(-\left(\frac{t}{\tau}\right)^{\alpha}\right),
	\end{equation}
	with the anomalous exponent $\alpha\in(0,1]$ and the time scale parameter $\tau>0$. This survival function can be writen as the Laplace transform of a fixed probability density function of the event rate $p(\lambda)$,i.e.,
	\begin{equation}\label{eq_LaplaceGillespie}
		\Psi(t)=\int_{\tau}^{\infty}\psi(\tau')d\tau'=\int_{0}^{\infty}p(\lambda)e^{-\lambda t}d\lambda.
	\end{equation}
	Differentiation of both sides of Eq.(\ref{eq_LaplaceGillespie}) yields the corresponding expression for the waiting time probability density,
	\begin{equation}\label{eq_LaplaceGillespie2}
		\psi(t)=\int_{0}^{\infty} p(\lambda)\lambda e^{-\lambda t}d\lambda.
	\end{equation}
	This shows that the Mittag-Leffler waiting time density can be interpreted as a mixture of infinitely many exponential waiting time densities $\lambda e^{-\lambda t}$, each with a $\lambda$-dependent weight determined by $p(\lambda)$. Since the survival function for 
	Mittag-Leffler distribution is completely monotonic, this ensures the existence of $p(\lambda)$. A closed analytical form of $p(\lambda)$ \cite{Gorenflo2002} is given by
	\begin{equation}
		p(\lambda)=\frac{1}{\pi}\frac{\tau^{-\alpha}\lambda^{\alpha-1}\sin(\alpha\pi)}{\lambda^{2\alpha}+2\tau^{-\alpha}\lambda^{\alpha}\cos(\alpha\pi)+\tau^{-2\alpha}}.
	\end{equation}
	The inverse transform method can then be applied to draw from this distribution as follows:
	\begin{equation}
		v=\int_0^\lambda\frac{1}{\pi}\frac{\tau^{-\alpha}\omega^{\alpha-1}\sin(\alpha\pi)}{\omega^{2\alpha}+2\tau^{-\alpha}\omega^{\alpha}\cos(\alpha\pi)+\tau^{-2\alpha}}d\omega,
	\end{equation}
	where $v$ is a uniform $[0,1]$ random variable. Substituting $u=\omega^{\alpha}$ and $du=\alpha\omega^{\alpha-1}d\omega$ and rearrange, we have
	\begin{equation}
		v=\frac{\tau^{-\alpha}\sin(\alpha\pi)}{\alpha\pi}\int_0^{\lambda^{\alpha}}\frac{1}{u^{2}+2\tau^{-\alpha}u\cos(\alpha\pi)+\tau^{-2\alpha}}du.
	\end{equation}
	Completing the square for $u$ in the denominator of the integrand
	\begin{equation*}
		v=\frac{\tau^{-\alpha}\sin(\alpha\pi)}{\alpha\pi}\int_0^{\lambda^{\alpha}}\frac{1}{(u+\tau^{-\alpha}\cos(\alpha\pi))^{2}+(\tau^{-\alpha}\sin(\alpha\pi))^{2}}du.
	\end{equation*}
	The integral can now be evaluated explicitly, which gives
	\begin{equation}
		\alpha\pi
		v=\tan^{-1}(\lambda^{\alpha}\tau^{\alpha}csc(\alpha\pi)+\cot(\alpha\pi))-\tan^{-1}(\cot(\alpha\pi)).
	\end{equation}
	This can be inverted by using the trigonometric identity $\tan(x-y)=(\tan(x)-\tan(y)/(1+\tan(x)\tan(y))$. After some manipulation one obtains
	\begin{equation}
		\lambda=\frac{1}{\tau}\left(\frac{\sin(\alpha\pi)}{\tan(\alpha\pi v)}-\cos(\alpha\pi)\right)^{-\frac{1}{\alpha}}.
	\end{equation}
	Finally, the random waiting time $\Delta t$ can be sampled from the exponential distribution with event rate $\lambda$, which gives
	\begin{equation}\label{eq_LaplaceGillespie3}
		\Delta t = -\frac{ln(u)}{\lambda} = -\tau \ln(u) \left(\frac{\sin(\alpha\pi)}{\tan(\alpha\pi v)}-\cos(\alpha\pi)\right)^{\frac{1}{\alpha}}.
	\end{equation}
	Here $u$ is another uniform $[0,1]$ random variable. (Note that the form given by Eq.(\ref{eq_LaplaceGillespie3}) is the same as that Kozubowski and Rachev form of Mittag-Leffler random variable derived independently in \cite{kozubowski1999univariate}).
	As a side note, we may also draw the minimum of an $\mathrm{Exp}[\mu]$ and a Mittag-Leffler$[\alpha,\tau]$ random variable as, 
	\begin{equation}
		y=-\frac{\ln(u)}{\mu+\frac{1}{\tau}\left(\frac{\sin(\alpha \pi)}{\tan(\alpha \pi v)}-\cos(\alpha \pi)\right)^{-\frac{1}{\alpha}}}.
	\end{equation}
	Sampling of the Mittag-Leffler waiting time through inverse transform method is particularly advantageous over the rejection sampling, as it gives rise to the explicit form of the waiting time density and there is no need for truncation of the power series in Eq.(\ref{eq_MLDefinition}) which is required for rejection sampling. Furthermore, the convergence of the Mittag-Leffler function is slow and usually requires the summation of a few hundred terms to achieve the desired accuracy, coefficient in each term also involves the gamma function which is computationally expensive, although a more efficient method based on the numerical inversion of the Laplace transform of the waiting time density has been introduced more recently \cite{GARRAPPAROBERTO2015NEOT}.
	
	\section*{Acknowledgements}
	This work was funded by Australian Research Council  grant number DP200100345.
	
	\bibliographystyle{unsrt}
	\bibliography{Stochastic_Compartment_Edit}
	
\end{document}